\documentclass[12pt]{article}
\usepackage{geometry}  
\usepackage{graphicx}				
\usepackage{amsmath}								
\usepackage{amssymb}
\usepackage{amscd}
\usepackage{yfonts}
\usepackage{amsmath,amsthm,amssymb}
\usepackage{amsthm}
\newtheorem{theorem}{Theorem}

              \newtheorem{lemma}[theorem]{Lemma}

\newcommand {\bp} {\bf p}
\newcommand {{\bx}} {\bf x}
\newcommand {{\bk}} {\bf k}
\newcommand {\bP} {{\bf  P}}

\date{}

\begin{document}
\medskip

\title{Scattering in  geometric approach to quantum theory. }
\author{Albert Schwarz}
\date{}							

\author {  A. Schwarz\\ Department of Mathematics\\ 
University of 
California \\ Davis, CA 95616, USA,\\ schwarz @math.ucdavis.edu}


\maketitle


 \begin{abstract}
 We define inclusive scattering matrix in the framework of geometric approach to  quantum field theory. We review the definitions of scattering theory in the algebraic approach and relate them to the definitions in geometric approach.
 \end {abstract}
  \section {Introduction}
  Geometric approach to quantum theory where the starting point is the set of states was suggested in \cite {GA1},\cite {GA}.  In this approach one can work with convex set  ${\cal C}_0$  of normalized states or with convex cone  $\cal C$ of not necessarily normalized states ( proportional points of the cone  $\cal C$ specify equivalent states) \footnote { We say that a closed convex set $\cal C$ is a convex cone if for every point $x\in \cal C$ all points of the form $\lambda x$ where $\lambda $ is positive also belong to $\cal C.$ Notice that in our terminology a vector space is a convex cone.}. In present paper we discuss scattering theory in geometric approach.    Our starting point is  a  convex cone $\cal C$ and a   subgroup $\cal V$ of the group of automorphisms of this cone.  
     
    We notice at the end of the paper that one can use also a subsemiring $\cal W$ of the semiring of endomorphisms $End {\cal C}.$ ( Endomorphisms of cone $\cal C$ form a semiring $End {\cal C}$ because the set of endomorphisms is closed with respect to addition and composition.
 Notice that the semiring $End {\cal C}$ is closed also with respect to multiplication by a nonnegative number; we assume that $\cal W$ also has this property.)
  
  We review  geometric and algebraic approaches to quantum theory and  the relation between these approaches. We  give  definitions of scattering matrix and inclusive scattering in algebraic approach.  This makes the present paper independent  of  papers  \cite {GA1},\cite {GA} and of the papers \cite {SC},  \cite {GA2} devoted to the scattering in algebraic approach.

Let  us recall the relation of the geometric approach  with  the algebraic approach to quantum theory \cite {GA}. In algebraic  approach a starting point is an associative algebra $\cal A$ with involution
 $ ^*$ ( a $*$-algebra). The cone ${\cal C}$  of not necessarily normalized states  is defined as a set of linear functionals on $\cal A$ obeying $f(A^*A)\geq 0.$ Every element $B\in\cal A$ specifies two operators on ${\cal A}^{\vee}$ ( on the dual space); one of them, denoted by the same symbol $B$, transforms a functional $f (A)$ into the functional $f(AB)$, another, denoted by the symbol $\tilde B$, transforms
$f(A)$  { into }  the functional $f(B^*A).$  The operator $\tilde B B$ is an endomorphism of the cone $\cal C.$   We define $\cal V$ as the group of all involution preserving automorphisms  of $\cal A$ acting in natural way on $\cal C.$ The semiring $\cal W$ is defined as the minimal set of endomorphisms of $\cal C$ containing all endomorphisms of the form $\tilde B B$ and closed with respect to addition and composition (it is closed also with respect to multiplication by a non-negative number as all semirings we consider).

 To define scattering in any approach to quantum field theory we need   notions of time and spatial translations.  In algebraic approach translations (as any symmetries) are automorphisms of the algebra $\cal A$; these automorphisms induce the automorphisms of the cone $\cal C$ and other objects related to the  algebra $\cal A.$  In geometric approach translations should be regarded as  elements of the group $\cal V$ consisting of
 automorphisms of the cone $\cal C$; their action on the cone should induce  an  action on the semiring $\cal W$.  
 
   Particles and  quasiparticles are defined as elementary excitations   of stationary translation-invariant state $\omega.$ 
   
   In algebraic approach one can define the notion of scattering  matrix of elementary excitations. Probably it is impossible to generalize this  notion  to  geometric approach, however,  in geometric  approach  one can give a  very natural definition of {\it inclusive} scattering matrix of elementary excitations of stationary translation-invariant state $\omega$ . It is easy to show that this notion agrees with analogous notion in algebraic approach.  
   
   Notice that our constructions can be applied also to scattering of quasiparticles in equilibrium and non-equilibrium statistical physics. ( The conventional scattering matrix does not make sense in this situation, but the inclusive scattering matrix does; see \cite {S},\cite {GA2}).
  
   In \cite {GA4} we apply the notions of present paper to define scattering in the framework of Jordan algebras.
 \section {Geometric approach}
 In geometric approach to quantum theory we  start with a  convex closed cone $\cal C$ of (non-normalized) states in Banach space $\cal L$ (or, more generally, in complete topological linear space $\cal L$).  We fix a subgroup $\cal V$  of the group  of automorphisms of the cone $\cal C.$ ( By definition an endomorphism of $\cal C$ is a continuous  linear operator in $\cal L$ transforming the cone into itself. An automorphism is an invertible endomorphism.)
 
  In  some cases it is useful to add to this data  a subsemiring $\cal W$ of  the semiring ${\rm End} (\cal C$) of endomorphisms of the cone; we assume that $\cal W$ is invariant with respect to the action of the group $\cal V$.

 The dynamics in quantum theory is governed by one-parameter group of  time translations $T_{\tau}$ acting on the cone $\cal C.$ We assume that $T_{\tau}\in \cal  V$. ( Here $\tau$ stands for a real number.) Time translations 
 can be considered also as  transformations of $\cal W$  denoted by the same symbol $T_{\tau}$. If 
 $A\in \cal V$ or $A\in \cal W$ the time translation acts as conjugation: $ T_{\tau}(A)= T_{\tau}AT_{-\tau}$;  we will use the  notation $T_{\tau}(A)=A(\tau).$ 

Quantum field theory in geometric approach is specified by a cone  $\cal C$ with the action of spatial translations $T_{\bx}$  where $\bx \in {\mathbb {R}}^d$ and time translations $T_{\tau}$ (the translations should constitute a commutative subgroup of the group $\cal V$.)  The same data specify statistical physics in the space ${\mathbb {R}}^d$ where $d$ stands for the dimension of the group of spatial translations. We use the notations 
$$T_{\tau}T_{\bx}(A)= T_{\tau}T_{\bx}AT_{-\tau}T_{-\bx}=A(\tau,{\bx})$$
for an operator $A$ acting in $\cal L.$

Let us discuss the relation of the above definitions to the  quantum theory in the algebraic approach. In this approach as in geometric one we need time and spatial translations to define  elementary excitations and scattering.  The time translations $T_{\tau}$ and spatial translations $T_{\bx}$ act as automorphisms of $\cal A$; these automorphisms induce  automorphisms of the cone $\cal C$ and of the semiring $\cal W$ denoted by the same symbols. If $\omega\in \cal C$ is a translation-invariant stationary state we can consider   a representation of $\cal A$ in a pre Hilbert space $\cal H$ such that there exists a cyclic vector $\theta\in \cal H$ obeying $\omega (A)=\langle \theta,A\theta\rangle.$ This representation is called GNS ( Gelfand-Naimark-Segal) representation. We denote an operator in this representation corresponding to $A\in \cal A$ by the same symbol $A.$ (Notice that these operators are bounded. )  We can consider also the representation of $\cal A$ in the Hilbert space $\bar {\cal H}$ (in the completion of $\cal H$). Time and spatial translations descend to  $\cal H$ and to $\bar {\cal H}.$

For every vector $\Psi$ in the Hilbert space $\bar{\cal H}$   we define the corresponding state $\sigma$ by the formula $\sigma (A)=\langle \Psi, A\Psi \rangle.$ If $\Psi=\theta$ we have $\sigma=\omega$, if $\Psi= B\theta$ we have $\sigma=\tilde B B\omega.$

\section {Elementary excitations}

Let us repeat the definitions and statements from \cite {GA} with small modifications.

We   consider a translation-invariant  stationary state $\omega\in \cal C.$  Let us start with the definition of  excitation of $\omega$ in geometric approach.  {\it  We  say that $\sigma \in \cal C$ is an excitation of $\omega$ if  $ T_{\bx}\sigma$ tends  to $C\omega $ as $\bx$ tends to $\infty$  for some constant $C.$}  (  We have in mind weak convergence in this definition. Recall  that  $u$  is a weak  limit of $u_{\alpha}\in \cal L$ if for every  $f\in {\cal L}^{\vee}$  (in the dual space)  the limit of 
$f(u_{\alpha}) $ is equal to $f(u)$.) We say that proportional elements of a cone specify the same state, hence this condition  means that for large $\bx$ the state $ T_{\bx}\sigma$ is close to $\omega.$

To define the notion of elementary excitation we need a notion of elementary space.

Recall that {\it elementary space} $\textgoth {h}$ is defined as a space of smooth real-valued  or complex-valued functions on $\mathbb{R}^d\times \cal I$ with all derivatives decreasing faster than any power (here $\cal I$ denotes a finite set consisting of $m$ elements).  One can identify this space with ${\cal S}^m$ (with direct sum of $m$ copies of Schwartz space ${\cal S}={\cal S}(\mathbb {R}^d$).  The space  $\textgoth {h}$ can be regarded as pre Hilbert space  (as  a  dense subspace of $L^2$). The spatial translations act naturally on   $\textgoth {h}$ ( shifting the argument); we assume that the time translations also act on $\textgoth {h}$ and commute with spatial translations. In momentum representation an element $\phi$ of $\textgoth {h}$ should  be considered as a complex function of ${\bk}\in \mathbb {R}^d$  and discrete variable $i\in \cal I$. If $\textgoth {h}$ consists of real-valued functions then in momentum representation we should impose the  condition $\bar\phi (-{\bk})=\phi (\bk).$ The spatial translation $T_{\bx}$ is represented as multiplication by $e^{i\bx\bk}$ and the time translation $T_{\tau}$ is represented as a multiplication by a matrix $e^{-i\tau E(\bk)} $ where $E(\bk)$ is a non-degenerate Hermitian matrix.  We assume that $E(\bk)$    a smooth function of at most polynomial growth, then the multiplication by $E(\bk)$ is an operator acting in $\textgoth {h}.$ The eigenvalues of   $E(\bk)$   are denoted by $ \epsilon_s ({{\bk}}).$

We need some facts about time evolution of elements of  $\textgoth {h}$ in coordinate representation.

If
$$  |(T_{\tau}\phi)({\bx}, j)|<  C_n (1+|{\bx}|^{2}+\tau^2)^{-n}$$
for  all   $\bx\in \mathbb {R}^d$ obeying  $ \frac {{\bx}}{\tau}\notin U$  and all $n\in \mathbb{N}$ we say that
$\tau U$  is an {\it  essential support} of   $T_{\tau}\phi $ in coordinate representation. Notice that the set $U$ is not defined uniquely; if $U'$ is a subset of  $ \mathbb {R}^d$ containing $U$ and  $\tau U$  is an essential support of   $T_{\tau}\phi $ in coordinate representation then $\tau U'$  is  also an essential support of   $T_{\tau}\phi $. 

Let us consider functions $f_1,...,f_n\in \textgoth {h}$ and essential supports $\tau U_{i}$
of functions $T_{\tau}(f_i)$ in coordinate representation.{\it  We say that these functions do not overlap if the distances between sets $U_{i}$ are positive  (the distances between essential supports grow linearly with $\tau$).}
\begin {quotation} \label {ASS}{\it   ASSUMPTION. We assume that collections   $ (f_1, ..., f_n)$ of non-overlapping functions are dense 
in ${\textgoth {h}}^n={\cal S}^{mn}$}
\end {quotation}

It is easy to verify that this assumption is almost always satisfied (in particular, it is satisfied if all functions $\epsilon_s ({{\bk}})$ are strictly convex). The proof can be based on the following lemma.

\begin {lemma} \label {K}
  Let us  denote by $U_{\phi}$ , where $\phi\in \textgoth {h}$,  an open subset of  $\mathbb {R}^d$ containing all points  having the form $ \nabla \epsilon_s ({{\bk}})$ where ${\bk}$ belongs to $\rm {supp}(\phi)=\cup_j \rm {supp}\phi _j)$ (to the union of supports of the functions $\phi (\bk,j)$ in momentum representation). 

Let us assume that   $\rm {supp}(\phi)$ is a compact subset of $\mathbb {R}^d$.
 Then  for large $|\tau|$ we have
$$ |(T_{\tau}\phi)({\bx}, j)|<  C_n (1+|{\bx}|^{2}+\tau^2)^{-n}$$
where $ \frac {{\bx}}{\tau}\notin U_{\phi}$, the initial data $\phi=\phi ({\bx},j)$ is the Fourier transform of $\phi ({\bk},j)$, 
and $n$ is an arbitrary integer. (In other words $\tau U_{\phi} $ is an essential support of  $T_{\tau}\phi$ in coordinate representation).
\end {lemma}
The proof of this lemma  ( Lemma 2 in \cite {GA}) can be given by means of the stationary phase method; see Section 4.2 of \cite {SC} for more detail.

{\it An elementary excitation of $\omega$ is defined as a map  $\sigma:\textgoth {h}\to \cal {C}$ of an elementary space  $\textgoth {h}$ into the set of excitations of $\omega$ . This map should  commute with translations  and satisfy the following additional requirement:  one can define  a map $L:\textgoth {h}\to End (\cal  L)$
obeying $\sigma(\phi)=L(\phi)\omega$.} 


Notice that the conditions we imposed on $L(\phi)$ do not specify it uniquely.
Later we impose some extra conditions  on these operators.
Not very precisely one can say that the operators $L(\phi)$ and $L(\psi)$ should almost commute  if
supports of $\phi$ and $\psi$ are far away  (see (\ref {COM}) 
 for precise formulation).  Still these extra conditions leave some freedom in the choice of $L$. We assume that the operators $L(\phi)$ are chosen in some way.

In algebraic approach we define an excitation of $\omega$ as a vector in the space of GNS representation $\cal H$; assuming cluster property one can verify that the state corresponding to such a vector is an excitation in the sense of geometric approach.  An elementary excitation  of $\omega$  is defined  as an isometric map  $\Phi$ of elementary space $\textgoth {h}$ into
 $\cal  H$ commuting with time and spatial translations. This definition agrees with the definition in geometric approach. To verify this fact we notice that the assumption that $\theta$ is a cyclic vector implies the existence of operators $B(\phi)$ obeying $\Phi (\phi)=B(\phi)\theta.$  (Here $\phi\in \textgoth {h}.$ ) We  define a   map $\sigma : \textgoth {h}\to \cal C$ saying that $\sigma(\phi)$ is a linear functional on $\cal A$ assigning  a 
number $\langle \Phi (\phi), A\Phi(\phi)\rangle$ to $A\in \cal A.$
The map $\sigma$ is quadratic if we are working over $\mathbb {R}$, it is Hermitian if we are working over $\mathbb {C}.$ It commutes with time and spatial translations. Representing
$\sigma(\phi)$ in the form $\sigma(\phi)= L(\phi)\omega$ where
$L(\phi)=\tilde B(\phi) B(\phi)\in End( \cal C)$ we obtain that this map specifies an elementary excitation in geometric approach.

{\it We assume that $B(\phi)$ is linear with respect to $\phi$;} then $L(\phi)$ is quadratic or Hermitian.

We say that a map $\sigma$ of real vector spaces is quadratic if the expression 
$\sigma (u+v)-\sigma (u) -\sigma (v)$ is linear with respect to $u$ and $v.$
A map $\sigma$ of complex vector spaces is Hermitian if 
$\sigma(u+v)-\sigma (u) -\sigma(v)$ is linear with respect to $u$ and antilinear with respect to $v.$ If $V$ is a real vector space then the corresponding cone $C(V)$ is defined  as a convex envelope of the set of vectors of the form $v\otimes v$ in the tensor square $V\otimes V$. (If we are dealing with topological vector spaces there exist different definitions  of tensor product and of topology in the tensor product. In this case  we should consider the closure of convex envelope in appropriate topology of tensor product. )   A quadratic map $V\to V'$ induces a linear map of the cone $C(V)\to V'$, a quadratic map of $V$  into a cone $C'\subset V'$ induces a linear map of cones $C(V)\to C'.$ Similar statements are true for complex vector spaces and Hermitian maps. ( The cone corresponding to complex vector space is defined as a convex  envelope of the set of vectors of the form $f\otimes \bar f$ in the tensor product  $V\otimes \bar V.$) If $V$ is a Hilbert space the corresponding cone can be identified with the cone of positive definite  self-adjoint operators belonging to trace class.


It is natural to assume that in geometric approach the maps  $\sigma$  and $L$ are  quadratic or Hermitian, but this assumption is not used in most of our statements.


Elementary excitations should be identified with particles or quasiparticles.  Notice that  particles and quasiparticles can be unstable; this means that we should consider also objects that only approximately obey the conditions we imposed on elementary excitations. The definition of inclusive scattering matrix given in the next section works also for such objects, but instead of the time $\tau$ tending to $\pm \infty$ we should consider large, but finite $\tau.$ (This is true also for the conventional scattering matrix in algebraic approach; see Appendix to \cite {SC} for detail.)

\section {Scattering. M\o ller matrices.}
Let us consider scattering of elementary excitations defined by the map $\sigma(f)=L(f)\omega.$ 

We define the operator $L(f,\tau)$ where $f\in \textgoth {h}$ by the formula
$$L(f,\tau)= T_{\tau}( L(T_{-\tau}f))= T_{\tau}L(T_{-\tau}f)T_{-\tau}.$$
(We are using the same notation for time translations in $\cal C$ and in $\textgoth {h}.$
The time translation acts on operators as conjugation with $T_{\tau}$.)
{ \it We assume that } $\sup _{\tau \in \mathbb {R}} ||T_{\tau}||<\infty$  {\it and the operators $L(f)$ are bounded, hence} $\sup _{\tau \in \mathbb {R}}||L(f,\tau)||<\infty.$  ( Here and in what follows we assume that $\cal L$ is a Banach space. If  $\cal L$ is a a topological vector space specified by a system of seminorms we should impose the above conditions for every seminorm.)

Notice  that $L(f,\tau)\omega$ { \it does not depend on } $\tau.$ (Using the fact that the map $\sigma$ commutes with translations we obtain that $L(f,\tau)\omega= T_{\tau} \sigma(T_{-\tau}f)=\sigma (f).$). This means that
\begin {equation} \label {DOT}
\dot L(f,\tau)\omega=0
\end {equation}
where  the dot stands for the derivative with respect to $\tau.$

Let us introduce the notation
\begin{equation}\label {INS}
\Lambda (f_1,\cdots,f_n|-\infty)= \lim _{\tau_1\to-\infty,\cdots, \tau_n\to-\infty}\Lambda (f_1, \tau_1,\cdots, f_n,\tau_n)
\end {equation}
where
$$ \Lambda (f_1,\tau_1,...,f_n,\tau_n)=L(f_1,\tau _1)...L(f_n,\tau_n)\omega.$$
We say that (\ref {INS}) is an $in$-state.

 For large negative $\tau$ the  state $$T_{\tau} \Lambda (f_1,\cdots,f_n|-\infty)$$ can be described  as a collection of  particles with wave functions $T_{\tau}f_i.$
To prove this fact we use the formulas
 $$T_{\tau}(L(f,\tau'))=T_{\tau+\tau'}L(T_{-\tau'}f) T_{-\tau-\tau'}=
 L(T_{\tau}f,\tau+\tau'),$$
   $$T_{\tau} \Lambda (f_1,\cdots,f_n|-\infty)=\Lambda (T_{\tau}f_1,\cdots ,T_{\tau}f_n|-\infty).$$ 
For $f_1, \cdots, f_n$ in a dense subset  of $\textgoth {h}\times \cdots \times \textgoth {h} $   the distance between essential supports of wave functions $T_{\tau}f_i$ tends to $\infty$  as
$\tau \to -\infty.$ This follows from  the assumption  in preceding section.

This remark allows us to say that for arbitrary $\tau$  the state $T_{\tau} \Lambda(f_1,\cdots,f_n|-\infty)$ describes a collision of particles with wave functions $(f_1,\cdots,f_n).$

It is obvious that {\it the $in$-state ( \ref {INS}) is symmetric with respect to 
$f_1,...,f_n$ if}   
\begin {equation} \label {SY}
\lim_{\tau \to -\infty} ||[ L(f_i,\tau), L(f_j,\tau)]||=0.
\end {equation}
 One can replace (\ref {SY}) by
\begin {equation}
\label {OM}
 ||[L(\phi),L(\psi)]||\leq \int d{\bx}d{\bx}' D^{ab}(\bx-\bx')|\phi_{\it a}(\bx)|\cdot |\psi_{\it b}(\bx')|
\end{equation}
where $D^{ab}(\bx)$ tends to zero faster than any power as $\bx\to \infty.$

Then the $in$-state is symmetric if the wave functions $f_1,...,f_n$  do not overlap.

Let us give conditions for the existence of the limit
\begin{equation}\label {LIM}
 \lim _{\tau_1\to-\infty,\cdots, \tau_n\to-\infty}\Lambda (f_1, \tau_1,\cdots, f_n,\tau_n).
\end{equation}

For simplicity we consider the case when  $\tau_1=\cdots=\tau_n=\tau.$
\begin {lemma} \label {1}
Let us  assume that  for $\tau \to -\infty$  the commutators   $[\dot L(f_i,\tau), L(f_j,\tau)]$ are small. More precisely, the norms of these commutators should be bounded from above by a summable function of $\tau:$  
\begin {equation}
\label {CC}
||[\dot L(f_i,\tau), L(f_j,\tau)]||\leq c(\tau), \\  \int|c(\tau)|d\tau<\infty.
\end{equation}  
 Then the vector $\Lambda (\tau)=
\Lambda (f_1, \tau,\cdots, f_n,\tau)$  has a limit as $\tau\to -\infty.$ 
\end {lemma}
It is sufficient to check that the 
norm  of the derivative of this vector with respect to $\tau$ is a summable function of $\tau$. (Then $\Lambda (\tau_2)-\Lambda (\tau_1)=\int_{\tau_1}^{\tau_2}\Lambda (\tau)d\tau$ tends to zero as $\tau_1,\tau_2\to-\infty.$)

Calculating $\dot \Lambda(\tau)$ by means of Leibniz rule we obtain $n$ summands; each summand has one factor with $\dot L$. 
The assumption about the behavior of commutators allows us to move the factor with derivative to the right if we neglect the terms tending to zero faster than a summable function of $\tau.$ It remains to notice that the expression with the derivative in the rightmost position vanishes due to (\ref {DOT}).

If $\cal L$ is a complete topological linear space with the topology specified by a system of seminorms we can  generalize the above  proof assuming an analog of (\ref {CC}) for  every seminorm.

Instead of (\ref {CC}) we can assume
that

\begin {equation}\label {CCD}
||[ L(f_i,\tau')-L(f_i,\tau),L(f_j,\tau)]||\leq c(\tau), \\  \int|c(\tau)|d\tau<\infty.
\end {equation}
where  $|\tau'-\tau|$ is bounded from above. 
 
 We can slightly strengthen (\ref {CC}) assuming that
 \begin {equation}
 \label {CDC}
||[\dot L(f_i,\tau), L(f_j,\tau_1)]||\leq c(\tau), \\  \int|c(\tau)|d\tau<\infty.
\end{equation}  
where $\tau-\tau_1$ is bounded from above. Then we can derive (\ref {CCD}) from (\ref {CDC}) integrating over $\tau$. 
 
 It is easy to derive from (\ref {CCD})  that 
\begin {equation}\label {CCDD}
||[ L(f_i,\tau')-L(f_i,\tau),L(f_j,\tau)]||\to 0
\end {equation}
as $\tau, \tau'\to \infty$ or $\tau, \tau'\to- \infty.$

\begin {lemma} \label {LLL} 
The condition (\ref{CCDD})  implies the existence of the  limit (\ref {INS}).
Hence the existence of this limit follows also from (\ref {CCD}) or (\ref {CDC}).
\end {lemma}
 We should check that the difference

$$L(f_1,\tau' _1)...L(f_n,\tau'_n)\omega -L(f_1,\tau _1)...L(f_n,\tau_n)\omega$$
tends to zero as $\tau'_i,\tau_i\to -\infty.$ 

It is sufficient to consider the expression
\begin {equation}\label {RR}
L(f_1,\tau _1)...( L(f_i,\tau'_i)-L(f_i,\tau_i))...L(f_n,\tau_n)\omega.
\end  {equation}
(One can go from $L(f_1,\tau _1)...L(f_n,\tau_n)\omega$ to $L(f_1,\tau' _1,...L(f_n,\tau'_n)\omega$ in $n$ steps changing one variable at every step.) Using (\ref {CCDD}) we can move  the factor $L(f_i,\tau'_i)-L(f_i,\tau_i)$ to the rightmost position in (\ref {RR}). It remains to notice that this factor gives zero acting on $\omega.$

Notice  that the distance between essential supports of functions $T_{\tau}f_i$ grows linearly as  $\tau \to -\infty$  if the sets $U_{f_i}$ do not overlap. This allows us to derive the existence of the limit  for  $f_1, \cdots, f_n$ in a dense  subset    of $\textgoth {h}\times \cdots \times \textgoth {h} $ 
 if we  assume that the commutator $[ T_{\alpha}(L(T_{-\tau'}f)),L(T_{-\tau}g)]$ is small when the essential supports of $T_{\tau'}f$ and $T_{\tau}g$ are far away for $\tau, \tau'\to \infty$. One can make this statement precise in various ways.

For example, applying Lemma \ref{LLL} we can prove the following theorem

\begin {theorem}\label{D}
 Let us assume that
\begin {equation}
\label {COM}
 ||[T_{\alpha}(L(\phi)),L(\psi)]||\leq \int d{\bx}d{\bx}' D^{ab}(\bx-\bx')|\phi_{\it a}(\bx)|\cdot |\psi_{\it b}(\bx')|
\end{equation}
where $D^{ab}(\bx)$ tends to zero faster than any power as $\bx\to \infty$ and $\alpha$ runs over a finite interval.
 Then the limit  (\ref {INS}) exists if the functions ${f_i}$ do not overlap (hence it exists  for  $f_1,...,f_n$ in dense  subset of $\textgoth {h}\times ...\times \textgoth {h}$).  
\end {theorem}

Applying (\ref {COM}) we obtain  estimates for  commutators 
$[T_{\alpha} (L(T_{-\tau}f)),L(T_{-\tau}g)]$  that are sufficient  to prove  the inequality    (\ref {CCD}), hence the existence of the limit (\ref {INS}).( We are using the relation
\begin {equation}\label{EQU}
||[L(f,\tau'),L(g,\tau)]||=||[T_{\tau'}(L(T_{-(\tau')}f), T_{\tau}(L(T_{-\tau}g))]||\leq
\end{equation}
$$C ||[T_{\tau'-\tau}(L(T_{-\tau'}f),L(T_{-\tau}g)]||$$
and its particular case for $\tau'=\tau.$)

Let us review shortly the scattering theory in the algebraic approach  modifying slightly the considerations of  \cite {GA2}\footnote {Notice that the operators $B(f,\tau)$ of present paper correspond to the operators $B(f\phi^{-1},\tau)$ of \cite {GA2}.  The  properties of operators $B(f,\tau)$ that are taken for granted  in the present paper are derived in \cite {GA2} from asymptotic commutativity of the algebra $\cal A$.}.  
Recall that in this approach an elementary excitation  of translation-invariant stationary state $\omega$ is specified by an isometric map
$\Phi:\textgoth {h}\to \cal H$ commuting with translations and obeying $\Phi (f)=B(f)\theta$ where $B(f)\in \cal A.$ ( Here $\theta$ stands for a a vector corresponding to $\omega$ in the space $\cal H$ of GNS representation.)

Let us define the operator $B(f, \tau)$ by the formula
$$ B(f,\tau)=T_{\tau}(B(T_{-\tau}f))=T_{\tau}B(T_{-\tau}f))T_{-\tau}.$$
Notice  that $B(f,\tau)\theta$ does not depend on $\tau$. This follows from the remark  that $\omega$ is stationary, hence 
$T_{-\tau}\theta=\theta$ and $B(f,\tau)\theta=T_{\tau} \Phi (T_{-\tau}f)=\Phi (f).$

\begin {lemma}\label {3}
Let us assume that 
$$||[ \dot B(f_i,\tau), B(f_j,\tau)]||\leq c(\tau)$$
where $c(\tau)$ is a summable function.  Then the  vector
$$\Psi (\tau)=B(f_1,\tau)...B(f_n,\tau)\theta$$
has a limit in $\bar {\cal H}$ as $\tau$ tends to $-\infty.$
\end {lemma}

\begin {theorem}\label {4}
Let us assume that
 \begin {equation}
\label {CM}
 ||[ \dot B(\phi),B(\psi)]||\leq \int d{\bx}d{\bx}' D^{ab}(\bx-\bx')|\phi_{\it a}(\bx)|\cdot |\psi_{\it b}(\bx')|
\end{equation}
{\it where $D^{ab}(\bx)$ tends to zero faster than any power as $\bx\to \infty.$
 Then  for  $f_1,...,f_n$ in dense  subset of $\textgoth {h}\times ...\times \textgoth {h}$  the  vector
$$\Psi (f_1,\tau_1, ..., f_n,\tau_n)= B(f_1, \tau_1) ...B(f_n,\tau_n)\theta$$ has a limit  in $\bar {\cal H}$ as $\tau_j$ tend to $-\infty$; this limit will be denoted by }
$$\Psi (f_1, ..., f_n| -\infty)$$
\end{theorem}
The proof of  Lemma \ref {3} is very similar to the proof of Lemma \ref {1}. To prove  Theorem  \ref {4}  we use the analog of (\ref {EQU})  to verify the analogs of (\ref {CDC}), (\ref {CCD}) and (\ref {CCDD}); using the analog of (\ref {CCDD}) we apply the method used in the proof of Lemma \ref {LLL}.

Let us introduce the asymptotic bosonic Fock space $\mathcal {H}_{as}$ as a Fock representation of 
canonical  commutation relations
$$[b(\rho), b(\rho')]=[b^+(\rho),b^+(\rho')]=0, [b(\rho),b^+(\rho')]=\langle \rho,\rho'\rangle$$
where $\rho, \rho'\in \textgoth {h}.$

We define M\o ller matrix $S_-$ as a linear map of ${\cal H}_{as}$ into 
$\bar  H$ that transforms $b^+(f_1) ...b^+(f_n)|0\rangle$ into
$\Psi (f_1, ..., f_n| -\infty)$.  ( Here $|0\rangle$ stands for the Fock vacuum.)
 Imposing some additional conditions one can prove that the operator $S_-$ can be extended to isometric embedding of ${\cal H}_{as}$ into $\bar H$ (see \cite {GA2}).   

Replacing $-\infty$ by $+\infty$ in the definition of $S_-$ we obtain the definition of the M\o ller matrix $S_+$. If both M\o ller matrices are surjective maps we say that the theory has particle interpretation.  We can define the scattering matrix  of elementary excitations (particles) as an operator in ${\cal H}_{as}$ by the formula $S=S_+^{*}S_-;$ if the theory has particle interpretation this operator is unitary.

 Let us define the $in$-operators $a^+_{in} $ by the formula
 \begin{equation}\label {IN}
a^+_{in}(f)= \lim_{\tau\to -\infty} B (f,\tau).
\end {equation}
This limit  exists as as strong limit on  vectors
 $\Psi (f_1, ..., f_n| -\infty)$ if there exists the limit  $\Psi (f,f_1, ..., f_n| -\infty).$
 
 Operators $a^+_{out} $ ($out$-operators)  are defined by the formula
 \begin{equation}\label {OUT}
a^+_{out}(f)= \lim_{\tau\to +\infty} B (f,\tau).
\end {equation}

Equivalently M\o ller matrix $S_-$  can be defined as a map $\mathcal {H}_{as}\to \overline {\cal H}$ obeying
$$a_{in}^+(\rho) S_-=S_-b^+(\rho), S_-|0\rangle=\theta.$$

The operators $a_{in}(\rho), a_{out}(\rho)$  (Hermitian conjugate to $a_{in}^+(\rho)$ and $a_{out}^+(\rho)$ )  obey
$$a_{in}(\rho) S_-=S_-b(\rho), a_{out}(\rho) S_+=S_+b(\rho).$$
Notice that spatial and time translations act naturally in   $\mathcal {H}_{as}.$  The M\o ller matrix  commutes with translations.

There exists an  obvious relation between our considerations in geometric and algebraic approach. It is clear that the operator 
$L(f,\tau)$ in the space of states corresponds to the operator $B(f, \tau)$ in $\bar {\cal H}$ (i.e $L(f,\tau)=\tilde B(f, \tau)B(f ,\tau).$) It follows that the state $\Lambda (f_1,\tau_1,\cdots, f_n,\tau_n)$ corresponds to vector $\Psi (f_1,\tau_1,\cdots, f_n,\tau_n)$ , the state
$\Lambda (f_1,\cdots,f_n|-\infty)$ (the $in$-state) corresponds to the  vector $\Psi (f_1,\cdots,f_n|-\infty).$

The relation  (\ref {COM}) implies that (\ref {LIM}) specifies a map of symmetric  power of $\textgoth {h}$ into the cone $\cal C$. This map  (defined on a dense subset) will be denoted by $\tilde S_-$; it can be regarded as an analog of the M\o ller matrix $S_-$ in geometric approach. The above statements allow us to relate $\tilde S_-$ with $S_-$  for theories that can be formulated algebraically. In this  case $S_-$ maps symmetric power of $\textgoth {h}$ considered as a subspace of the Fock space  into $\bar {\cal H}.$ Composing this map with the natural map of $\bar {\cal H}$ into the cone of states $\cal C$ we obtain $\tilde S_-.$

 The map $\tilde S_-$ is not linear, but in the case when $L$ is quadratic or Hermitian it  induces a multilinear map of the symmetric power  of the cone $C(\textgoth {h})$ corresponding to $\textgoth {h}$ into the cone $\cal C.$ 
 

 
 Constructing the scattering matrix in algebraic approach we imposed some conditions on commutators  ( for example the condition (\ref {CM})  in Lemma 5). These conditions can be replaced  by similar conditions on anticommutators, the above statements remain correct after slight modifications. ( In particular, we should consider the fermionic  Fock space instead of bosonic one.) It is important to notice that operators $L=\tilde B B$ (almost) commute not only in the case when operators $B$  (almost) commute, but also in the case when operators $B$ (almost) anticommute, hence our considerations in geometric approach can be applied not only to bosons, but also to fermions.
 
\section { Inclusive scattering matrix}
Instead of the cone $\cal C$ one can consider the dual cone ${\cal C}^{\vee}\subset {\cal L}^{\vee}$ (it consists of linear functionals that are non-negative on $\cal C$). The group $\cal V$ (in particular  the group of translations) and the semiring $\cal W$  act on ${\cal C}^{\vee}.$

Let us consider a  translation invariant stationary  element $\alpha\in {\cal C}^{\vee}$ obeying the conditions similar to the conditions we imposed on $\omega.$  ( In algebraic approach we can take $\alpha (\sigma)=\sigma (1)$, the value of $\sigma$ on the unit of algebra.) Let us assume that $\langle\alpha| L'(g)$  is an elementary excitation of $\alpha.$
 ( Here $L'$ maps the elementary space $\textgoth {h}$ into the space of  endomorphisms of  $\cal L$;  these endomorphisms can be considered also as endomorphisms of the dual space ${\cal L}^{\vee}$.)

Taking 
$$\lim_{\tau_k\to +\infty}\langle \alpha|(L'(g_1,\tau_1)...L'(g_m, \tau_m) |\Lambda(f_1,\cdots,f_n|-\infty)\rangle$$  we obtain a number characterizing the result of the collision.  We can write this number as
\begin {equation}\label {BBK}
\lim _{\tau_k'\to+\infty,\tau_j\to -\infty}\langle \alpha|L'(g_1,\tau'_1)...L'(g_m, \tau'_m) L(f_1,\tau_1)...L(f_n, \tau_n)|\omega\rangle
\end{equation}
Let us assume that {\it operators $L(f)$ obey (\ref {COM}) and operators $L'(g)$ obey similar condition}. Then
\begin{theorem}\label {INCL}
If  both $(f_1,..,f_n)$ and $(g_1, ..., g_m)$ do not overlap the limit (\ref {BBK}) exists. This limit  is symmetric with respect to  $(f_1,..,f_n)$ and  with respect to $(g_1, ..., g_m)$
\end {theorem}

The proof of this theorem  is similar to the proof of Theorem \ref {D}. The second statement  follows from the fact that operators $L(f_j,\tau_j)$ and $L(f_{j'},\tau_{j'})$  almost commute in the limit $\tau_j, \tau_{j'}\to -\infty$ and from similar fact
for operators $L'.$

By definition of elementary excitation $\sigma (\phi)$  is a quadratic  (or Hermitian) map, hence it is natural to assume that the map $L(\phi)$ is also quadratic (or Hermitian). Then it can be extended to a bilinear (or sesquilinear) map
$L(\tilde \phi, \phi)$ and the map $L\phi,\tau)$ can be extended to a map $L(\tilde \phi, \phi, \tau).$ (If we assume that the bilinear map is symmetric then these extensions are unique, but in  algebraic approach it  is convenient to consider  extensions that are  not symmetric. Recall that in the algebraic  approach we define $ L(\phi)$ as $\tilde B (\phi) B(\phi)$; the extension can be defined by the formula $L(\tilde \phi, \phi)=\tilde B(\tilde \phi)B(\phi).$) We assume that $L'$ is also quadratic or Hermitian and extend it to  bilinear or sesquilinear map.

Using these extensions we can define a functional
$$\sigma (\tilde g'_1,g'_1,..., \tilde g'_{n'}, g'_{n'},\tilde g_1,g_1,...,\tilde g_n, g_n)=$$
\begin {equation} \label {BKK}
\langle \alpha |  \lim _{\tau'_i\to +\infty, \tau_j\to -\infty} L'(\tilde g'_1,g'_1,\tau' _1)...L'(\tilde g'_{n'}, g'_{n'},\tau'_{n'})L(\tilde g_1,g_1,\tau _1)...L(\tilde g_n,g_n,\tau_n)|\omega\rangle
\end {equation}
that is linear or antilinear with respect to all of its arguments.

Notice that in the case when we take symmetric extensions of $L$ and $L'$ the existence of the limit (\ref {BKK}) follows from the existence of the limit (\ref {BBK});
in general case we should modify slightly the condition (\ref {COM}) to prove a generalization of Theorem \ref {INCL}.

We say that (\ref {BKK}) is  inclusive scattering matrix. ( If we do not assume that the map $L(\phi)$ is quadratic or Hermitian the inclusive scattering matrix should be defined by the formula (\ref{BBK}).) \footnote {Notice, that (\ref{BBK}) and (\ref {BKK}) can be considered either as inclusive scattering matrix of elementary excitations of state $\omega$ or as  inclusive scattering matrix of elementary excitations of state $\alpha$. Similar statement is true for analogs of Green functions introduced in Section 6. It is not clear whether this strange duality has any physical meaning.} This terminology comes from the fact that in algebraic approach  matrix elements of inclusive scattering matrix are related to inclusive cross-sections. In this approach one can express
 inclusive scattering matrix  in terms of on-shell GGreen functions that appear in the formalism of $L$-functionals (used in \cite {S}, \cite {SC}, \cite {MO}) and in Keldysh formalism \cite {UNI},\cite {CU},]\cite {K}.  Let us sketch the derivation of this expression.( See  \cite {S},\cite {SC},\cite {MO} for more detail.)

The functional  (\ref {BKK}) can be considered as a generalized function
\begin{equation} \label {ME}
\sigma(\tilde{\bk}'_1, \tilde i'_1, {\bk}'_1,i'_1,...,\tilde{\bk}'_{n'},\tilde i'_{n'}, {\bk}'_{n'}, i'_{n'}, \tilde{\bk}_1, \tilde i_1, {\bk}_1,i_1,..., \tilde{\bk}_{n},\tilde i_n, {\bk}_{n}, i_n)
\end {equation}
This generalized function is defined for open dense subset of its arguments.
It is sufficient to require that $\tilde{\bk}'_i\neq \tilde{\bk}'_j, {\bk}'_i\neq {\bk}'_j, \tilde{\bk}_i\neq \tilde{\bk}_j, {\bk}_i\neq {\bk}_j,$ for $i\neq j $ if we assume that ${\bk}\neq {\bk}'$ implies  $\nabla \epsilon_j({\bk})\neq \nabla \epsilon_j' ({\bk}')$ ( Recall that  we use the notation $ \epsilon _j({\bk})$ for eigenvalues of the matrix $E(\bk)$.) More generally we can consider the sets $U(\bk )$ consisting of  vectors $\nabla \epsilon _j({\bk})$  and assume that the sets $U(\bk)$ and $U(\bk')$ do not overlap. Then  the essential support of a function $T_{-\tau}(f)$  is far away from the essential support of a function $T_{-\tau}(f')$ if the support of $f$ lies in the neighborhood of $\bk$, the support of  $f'$ lies in the neighborhood of ${\bk}'\neq {\bk}$  and $\tau\to\infty.$

One can say that  the function (\ref {ME}) gives matrix elements of inclusive scattering matrix.  

Let us show that in the algebraic approach inclusive cross-sections can be expressed in terms of these matrix elements.
Notice that in this approach
\begin {equation} \label {AKK}
\begin{split}
\sigma (\tilde g'_1,g'_1,..., \tilde g'_{n'}, g'_{n'},\tilde g_1,g_1,...,\tilde g_n, g_n)=
\langle 1 |  \lim _{\tau'_i\to +\infty, \tau_j\to -\infty} \\ \tilde B'(\tilde g'_1,\tau'_1) B'(g'_1,\tau' _1)...\tilde B'(\tilde g'_{n'},\tau'_n)B'( g'_{n'},\tau'_{n'})\tilde B(\tilde g_1,\tau_1) B(g_1,\tau _1)...\tilde B (\tilde g_n,\tau_n)
B (g_n,\tau_n)|\omega\rangle=\\ \langle a^+_{out}(\tilde g'_1)...a^+_{out}(\tilde g'_{n'})\Psi (\tilde g_1,...,\tilde g_n|-\infty),a^+_{out}(g'_1)...a^+_{out}(g'_{n'}) \Psi(g_1,...,g_n|-\infty)\rangle=\\
\langle a_{out}(g'_{n'}) ...,a_{out}(g'_1)a^+_{out}(\tilde g'_1)...a^+_{out}(\tilde g'_{n'})\Psi (\tilde g_1,...,\tilde g_n|-\infty), \Psi(g_1,...,g_n|-\infty)\rangle
\end{split}
\end {equation}
We have used  {Theorem \ref{4},Eq.}(\ref {OUT}) and relations $(\tilde B_1B_2\omega)(A)= \omega(B^*_1AB_2)=\langle \theta,, B^*_1AB_2\theta\rangle=\langle B_1\theta, AB_2\theta\rangle$, $\langle 1|\tilde B_1B_2|\omega\rangle= \langle B_1\theta,B_2\theta\rangle$ in this derivation.

In terms of generalized functions
\begin{equation} \label {MEE}
\sigma (\tilde{\bk}'_1, \tilde i'_1, {\bk}'_1,i'_1,...,\tilde{\bk}'_{n'},\tilde i'_{n'}, {\bk}'_{n'}, i'_{n'}, \tilde{\bk}_1, \tilde i_1, {\bk}_1,i_1,..., \tilde{\bk}_{n},\tilde i_n, {\bk}_{n}, i_n)=
\end {equation}
$$\langle a_{out}({\bk}'_{n'}, i'_{n'}) ...a_{out}({\bk}'_1,i'_1)a^+_{out}(\tilde{\bk}'_1, \tilde i'_1)...a^+_{out}(\tilde{\bk}'_{n'},\tilde i'_{n'})\Psi (\tilde{\bk}_1, \tilde i_1, ...,\tilde{\bk}_{n},\tilde i_n)|-\infty), \Psi({\bk}_1,i_1,..., {\bk}_{n}, i_n|-\infty)\rangle$$

Inclusive scattering matrix can be expressed in terms of generalized Green functions.   { These functions
(GGreen functions) are defined by the formula}
\begin {equation} \label {GKK}
\langle 1 | T(\tilde B'(\tilde g'_1,\tilde\tau'_1) B'(g'_1,\tau' _1)...\tilde B'(\tilde g'_{n'},\tilde\tau'_n)B'( g'_{n'},\tau'_{n'})\tilde B(\tilde g_1,\tilde\tau_1) B(g_1,\tau _1)...\tilde B (\tilde g_n,\tilde \tau_n)B (g_n,\tau_n))|\omega\rangle
\end {equation}
where $T$ stands for chronological product (see \cite {GA2}).

The inclusive cross-section  of the process $(M,N)\to (Q_1 ...,Q_m)$ is defined as a sum (more precisely a sum of integrals)  of effective cross-sections  of the processes $(M,N)\to (Q_1,...,Q_m, R_1, ..., R_n)$ over all possible $R_1,...,R_n.$ If the theory does not have particle interpretation this  formal definition of inclusive cross-section does not work, but still the inclusive cross-section can be defined in terms of  probability of the process
$(M,N)\to (Q_1,...,Q_n+$ something else) and expressed in terms of inclusive scattering matrix defined above. To verify this statement we
consider the expectation value
\begin{equation}
 \label{eq:NU}
 \nu (a^+_{out}({\bp}_1,k_1)a_{out}({\bp}_1,k_1)\dots a^+_{out}({\bp}_m, k_m)a_{out}({\bp}_m,k_m))
\end{equation}
where $\nu $ is an arbitrary state.

This quantity is the probability density in momentum space for finding $m$ outgoing particles of the types $k_1,\dots ,k_n$ with momenta $\bp_1,\dots ,\bp_m$ plus other unspecified outgoing particles.
It gives inclusive cross-section if $\nu$ is an $in$-state.

Comparing this statement with  (\ref {MEE}) we obtain that inclusive cross-section can be obtained from inclusive scattering matrix if $\tilde {\bk}_i$ tends to  ${\bk}_i$
and $\tilde {\bk}'_i$ tends to ${\bk}'_i.$  (We  assume that  the expression
\begin{equation}
 \label{eq:NUU}
 \nu (a^+_{out}({\tilde\bp}_1,k_1)a_{out}({\bp}_1,k_1)\dots a^+_{out}({\tilde\bp}_m, k_m)a_{out}({\bp}_m,k_m))
\end{equation}
tends to (\ref {eq:NU}) as $\tilde {\bp}_i$ tends to $\bp_i.$)
\section {Analogs of Green functions}
Let us consider a functional
\begin {equation} \label {BKKL}
\langle \alpha |  T( L'(\tilde g'_1,g'_1,\tau' _1)...L'(\tilde g'_{n'}, g'_{n'},\tau'_{n'})L(\tilde g_1,g_1,\tau _1)...L(\tilde g_n,g_n,\tau_n))|\omega\rangle
\end {equation}
where $T$ denotes chronological product.
This expression is linear or antilinear with respect to  its arguments $g'_i,g_j.$  We assume that these  arguments  do  non-overlap. It follows from this assumption and  the  second statement of Theorem \ref {INCL}  (or generalization of this theorem) that  (\ref {BKKL}) tends to inclusive scattering matrix (\ref {BKK}) as 
$\tau'_k\to +\infty, \tau _j\to -\infty$  (the time ordering is irrelevant for the  first $n'$ factors and also for the last $n$ factors).

The functional  (\ref {BKKL}) can be considered as a generalized function
\begin{equation} \label {MEL}
G_{n',n} (\tilde{\bk}'_1, \tilde i'_1, {\bk}'_1,i'_1,\tau'_1...,\tilde{\bk}'_{n'},\tilde i'_{n'}, {\bk}'_{n'}, i'_{n'},\tau'_{n'}, \tilde{\bk}_1, \tilde i_1, {\bk}_1,i_1,\tau_1,..., \tilde{\bk}_{n},\tilde i_n, {\bk}_{n}, i_n,\tau_n)
\end {equation}
This generalized function is defined for open dense subset of its arguments.

One can obtain  (\ref {ME}) ( matrix elements of inclusive scattering matrix ) from (\ref {MEL}) taking the limit $\tau'_k\to +\infty, \tau _j\to -\infty.$

The function (\ref {MEL})  can be considered as an analog of Green function in $(\bp,t)$-representation.  Taking Fourier transform with respect to  $\tau'_k, \tau _j$ we obtain an analog of Green function in $(\bp, \omega)$-representation that also can be used to  calculate matrix elements of inclusive scattering matrix. ( If a function $f(t)$ has limits as $t\to \pm\infty$ then these limits can be calculated as residues in the poles of the  Fourier transform of $f(t)$).

In the algebraic approach the functional (\ref {BKKL}) and generalized  function (\ref {MEL}) are related to generalized Green function (GGreen function) \cite {GA2}.  Namely, in this approach one can obtain (\ref {BKKL}) from (\ref {GKK}) taking $\tilde \tau'_k=\tau'_k,
\tilde \tau_j=\tau_j $ and using the relation
$L(\tilde g,g,\tau)=\tilde B(\tilde g, \tau) B(g,\tau).$


\section {Discussion}

Let us discuss some properties of the above  construction of $in$-state  and of inclusive scattering matrix.

We start again with elementary excitation $\sigma:\textgoth {h}\to \cal C$ of state $\omega.$ By definition of elementary excitation there exists a map $L:\textgoth {h}\to End (\cal L)$ obeying $\sigma (\phi)=L(\phi)\omega.$ The map $L$ is not unique; let us prove that under some conditions the $in$-state does not change when we are changing $L.$ More precisely we can prove the following statement:

{\it  Let us assume that the maps $L_i: \textgoth {h}\to End (\cal L)$ can be used to define $in$-state and
$$ ||[L_i(\phi),L_j(\psi)]||\leq \int d{\bx}d{\bx}' D^{ab}(\bx-\bx')|\phi_{\it a}(\bx)|\cdot |\psi_{\it b}(\bx')|.
$$
where $D^{ab}$ tends to zero faster than any power. Then}
$$\Lambda (f_1,\cdots,f_n|-\infty)= \lim _{\tau_1\to-\infty,\cdots, \tau_n\to-\infty}L_{i_1(}f_1,\tau _1),...L_{i_n}(f_n,\tau_n)\omega.$$
({\it We assume that the functions ${f_i}$ do not overlap.)}

To  prove this statement we notice first of all that $L_i(f,\tau)\omega=L_j(f,\tau)\omega$ hence the choice of the operator $L_i$ in the rightmost position does not matter. Then we use the fact that  one can move every factor to the rightmost position without changing the limit (the commutators are small when $\tau_j\to -\infty.$)

Similar statement is true for inclusive scattering matrix.

Let us consider a Poincar\'e -invariant theory. 
Recall that in our definitions we started with the homomorphism of the translation group  $\cal T$ into group of $\cal V$. We assume that this homomorphism can be extended to a homomorphism of the Poincar\'e group $\cal P.$  The translation group acts also on the elementary space $\textgoth {h}$; we assume that this action also can be extended to the action of Poincar\'e group and that the  elementary excitation  of Poincar\'e invariant state $\omega$ considered as a map $\sigma:\textgoth {h}\to \cal C$ commutes with the actions of Poincar\'e group on $\textgoth {h}$ and $\cal C:$ for every $P\in\cal P$ and $f\in \textgoth {h}$ we have
\begin {equation} \label {SIG}
\sigma (Pf)=P\sigma(f)
\end{equation}
Then we say that the theory is Poincar\'e-invariant. 

By definition of elementary excitation there  exists a map $L:\textgoth {h}\to \cal W$ obeying $\sigma (f)=L(f)\omega.$ If $L$ commutes with Poincar\'e transformations 
 the scattering is  obviously Poincar\'e-invariant.  However, one can prove Poincar\'e invariance of scattering in much more general situation. Let us sketch a proof of this fact assuming that
\begin{equation}\label {IND}
\lim_{\tau\to-\infty}||[L(Pf_i, \tau), L^P(f_j,\tau)||=0
\end {equation}
(We introduced notation $L^P(f,\tau)=PL(f,\tau)P^{-1}.$)

The generalized M\o ller matrix $\tilde S_-$  is a map of the symmetric power of $\textgoth {h}$ into $\cal C$. Let us check that this map commutes with actions of Poincar\'e group. (Similar proof can be applied to inclusive scattering matrix.)

We should identify
\begin {equation} \label {LOR}
 L(Pf_1,\tau ),...L(Pf_n,\tau)\omega
 \end {equation}
with
$$PL(f_1,\tau ),...L(f_n,\tau)\omega=L^P(f_1,\tau ),...L^P(f_n,\tau)\omega$$
in the limit $\tau\to-\infty.$ 
We will show that we can replace $L(Pf_i, \tau)$ with $L^P(f_i,\tau)$ in any number of factors of (\ref {LOR})  without changing the limit.  For  the rightmost factor this statement is equivalent to (\ref {SIG}).  Let us assume that this statement is correct for the last $k$ factors. Then it is true also for $(k+1)$-th factor from the right. (To prove this we interchange the $(k+1)$-th factor with $k$-th factor from the right   using (\ref {IND}) and use the induction hypothesis.)  We proved the statement by induction.

Modifying the   considerations of  Section 4 we can give various conditions for Poincar\'e invariance of scattering thery on a dense subset of
$\textgoth {h}\times ...\times \textgoth {h}.$
\vskip .1in
Until now we did not use the semiring $\cal W$ in our considerations.  Let us show  how it can be used. We need an additional structure on this semiring: we assume that it is represented as a union of subsemirings ${\cal W}_V$ corresponding to domains $V\subset \mathbb {R}^d.$   If $L_1\in {\cal W}_{V_{1}}$, $L_2\in {\cal W}_{V_{2}}$, $||L_1||=||L_2||=1$ and the domains $
$ are far away we assume
that the commutator $[L_1,L_2]$ is small: for every $n$
$$||[L_1,L_2]||\leq C_n d(V_1, V_2)^{-n}$$
where $d(V_1, V_2)$ stands for the distance between domains and $C_n$ is a constant factor.

Let us assume  that the operators $L(\phi)$ belong to the semiring $\cal W$. Moreover, we require that in the case when  the function 
$T_{\tau} \phi$ has essential support in $\tau V$ the corresponding operator 
$L(T_{\tau} \phi)$ belongs to ${\cal W}_{C\tau V}$ for some constant $C.$
Then  it is easy to check that the inequality (\ref {CCD}) is satisfied in the case when functions $f_i,f_j$ do not overlap. This allows us to prove the existence of the limit (\ref {INS}) defining $in$-state in the case when the functions $f_i$ do not overlap.

One can give a formulation of quantum theory in terms of group $\cal V$ of linear operators acting in topological vector space $\cal$ and semiring $\cal W$ of linear operators  { acting} in the same space. It seems that such a formulation can be useful in $BRST$ approach to quantum theory.

 { One can prove  analogs of results of present paper in the case when the group of spatial translations is discrete.  It is natural to assume that this group is is isomorphic to  $\mathbb{Z}^d$ (free abelian group with $d$ generators). This happens, in particular, for quantum theory on a lattice in $d$-dimensional space.

The notion of elementary space should be modified: $\textgoth {h}$ should consist of  fast decreasing functions on the lattice $\mathbb{Z}^d$, spatial translations act on this space as  shifts of the argument. Equivalently one can consider elements of $\textgoth {h}$ as smooth functions on  a torus (as smooth periodic functions of $d$ arguments); taking corresponding Fourier series we come to fast decreasing functions on $\mathbb{Z}^d$.

Working with this version of elementary space we can modify all definitions and theorems
of this paper. One should expect that  modified theorems can be applied to gapped lattice systems.

These ideas can be applied also in the case when translation symmetry is spontaneously broken (i.e.the theory is translation invariant, but we consider elementary excitations of a state $\omega$ that is invariant only with respect to a discrete subgroup of the translation group.).

Similar modifications  can be made when the time is discrete.}

{\bf Acknowledgements} I am indebted to A. Rosly for very useful comments.




\begin{thebibliography}{10}
 
\bibitem {GA1} Schwarz A. Geometric approach to quantum theory. SIGMA. Symmetry, Integrability and Geometry: Methods and Applications. 2020 Apr 1;16:020.
\bibitem {GA} Schwarz, A., 2021. Geometric and algebraic approaches to quantum theory. Nuclear Physics B, 973, p.115601.


\bibitem {GA2}Schwarz, A., 2021. Scattering in algebraic approach to quantum theory. Associative algebras. arXiv preprint arXiv:2107.08553.

\bibitem {GA4} Schwarz A. Scattering  in algebraic approach to quantum theory . Jordan algebras (in preparation)

\bibitem {S} Schwarz A. Inclusive scattering matrix and scattering of quasiparticles. Nuclear Physics B. 2020 Jan 1;950:114869.
\bibitem {SC} Schwarz, A., 2019. Scattering matrix and inclusive scattering matrix in algebraic quantum field theory. arXiv preprint arXiv:1908.09388.
\bibitem{T} Tyupkin, Yu, On the adiabatic definition of the S matrix in the formalism of L-functionals,  Theoretical and Mathematical Physics, 1973, 16:2, 751-756, https://link.springer.com/content/pdf/10.1007\%2FBF01037126.pdf
\bibitem {UNI}Chou, K.C., Su, Z.B., Hao, B.L. and Yu, L., 1985. Equilibrium and nonequilibrium formalisms made unified. Physics Reports, 118(1-2), pp.1-131.
\bibitem {CU} Chu, H., and H. Umezawa. A unified formalism of thermal quantum field theory. International Journal of Modern Physics A 9.14 (1994): 2363-2409.
\bibitem {K}   Kamenev, Alex, and Alex Levchenko. Keldysh technique and non-linear  sigma-model: basic principles and applications. Advances in Physics (2009).
\bibitem {MO} A. Schwarz, Mathematical foundations of quantum field theory, World Scientific 
\end{thebibliography}
\end {document}